\documentclass[aps,twocolumn,prd,superscriptaddress,showpacs,preprintnumbers,nofootinbib]{revtex4-1}

\usepackage{color}
\usepackage{graphicx}
\usepackage{amsmath, amssymb}
\usepackage{amsfonts}
\usepackage{dcolumn}
\usepackage{subfigure}
\usepackage{dsfont}

\newcommand{\Tr}{\ensuremath{\operatorname{Tr}}}

\def\eq#1{(\ref{#1})}
\def\Eq#1{Eq.~(\ref{#1})}
\def\eqref#1{(\ref{#1})}

\def\lA0{{\langle A_0 \rangle}}
\def\bA0{{\bar{A}_0}}

\def\0#1#2{\frac{#1}{#2}}

\graphicspath{{../figures/}{./}}



\newcommand{\Slash}[1]{\ooalign{\hfil/\hfil\crcr$#1$}}
\newcommand{\pa}{\partial}
\newcommand{\nn}{\nonumber}

\newcommand{\ep}{\epsilon}

\newcommand{\al}{\alpha}
\def\pa{\partial}
\def\al{\alpha}

\def\dr{{D\!\llap{/}}\,}

\begin{document}

\author{Yuji Igarashi}
\affiliation{Faculty of Education, Niigata University, Niigata 950-2181, Japan}
\author{Katsumi Itoh}
\affiliation{Faculty of Education, Niigata University, Niigata 950-2181, Japan}
\author{Jan M. Pawlowski}
\affiliation{Institut f\"ur Theoretische Physik, Universit\"at Heidelberg, 
Philosophenweg 16, D-69120 Heidelberg, Germany}
\affiliation{ExtreMe Matter Institute EMMI, GSI Helmholtzzentrum f\"ur 
Schwerionenforschung mbH, Planckstra{\ss}e 1, D-64291 Darmstadt, Germany}
\title{Functional flows in QED and the modified Ward-Takahashi identity}
\begin{abstract}
 In the functional renormalisation group approach to gauge theory,
  the Ward-Takahashi identity is modified due to the presence of an
  infrared cutoff term. It take the most accessible form
  for the Wilsonian effective action. 

  In the present work we solve these identities, partially, for the
  Wilson effective action of QED. In particular, we compute the
  longitudinal part of the photon two point vertex function as a
  momentum-dependent function in the presence of the cutoff $k$. The
  resultant Wilsonian effective action carries form factors that
  originate from the modified Ward-Takahashi identity. We show how
  this result carries over to the one-particle-irreducible effective
  action. 
\end{abstract}

\maketitle

\section{Introduction}

The functional renormalisation group (FRG) approach,
\cite{Wilson:1971bg,Wilson:1971dh,Wegner:1972ih,Polchinski:1983gv,Wetterich:1992yh}
has been successfully applied to various
field theoretical problems as a robust non-perturbative method.
Applications range from quantum gravity, high energy physics and QCD
to problems in condensed matter physics, and non-equilibrium physics.

The FRG approach is based on a flow equation for a generating
functional of the theory at hand, and hence can be formulated in terms
of a coupled set of integro-differential equations for correlation
functions. The full, field-dependent propagator takes a pivotal
r$\hat{\rm o}$le in these formulations. In gauge theories, such a
set-up requires a gauge fixing, and gauge invariance is carried by the
Slavnov-Taylor (STI) or Ward-Takahashi identity (WTI), henceforth both
summarised as WTI.  In the presence of an infrared momentum cutoff $k$
the WTI survives as modified Ward-Takahashi (mWTI) identity,
\cite{Bonini:1993kt,Ellwanger:1994iz,Bonini:1994kp,Freire:1996db,
  Becchi:1996an,D'Attanasio:1996jd,Reuter:1997gx,Igarashi:2001mf,
  Freire:2000mn,Pawlowski:2005xe,Sonoda:2007dj,Igarashi:2008bb,Igarashi:2009tj}.
The standard WTI is recovered in the limit of vanishing infrared
cutoff.

Still, even in the presence of the regulator the mWTI can be
formulated as standard WTI with a modified generator of gauge
transformations,
\cite{Igarashi:2001mf,Sonoda:2007dj,Igarashi:2008bb,Igarashi:2009tj}.
For more details on the properties of such a modified generator see
\cite{Bergner:2012nu}. This concise form of the mWTI as a
symmetry identity is very useful for the construction of closed
solutions of the mWTI. In the present work we use it for the
Wilson effective action $S_{k}$, the generating functional for
amputated connected correlation functions.

In summary, the FRG approach to gauge theories is based on master
equations for the scale dependence of the generating functionals, the
functional flow equations, as well as master equations for the
symmetries, the mWTI. In terms of the Wilsonian effective action,
$S_k$, both master equations are finite sums of linear and bilinear
forms in $S_k^{(n)}$, reading
\begin{align}\label{eq:MasterSLambda}
  D[\phi] S_k[\phi] = \sum c_{n} S_k^{(n)}[\phi]+\sum c_{nm} S_k^{(n)}[\phi]\,
  S_k^{(m)}[\phi] \,,
\end{align}
with operator coefficients $c_n$ and $c_{nm}$. For the flow equation of $S_k$, 
\cite{Polchinski:1983gv,Keller:1990ej,Keller:1991cj}, $D[\phi]$ stands
for the scale derivative $D[\phi] =k\partial_k$. In the case of the
mWTI, $D[\phi]$ stands for the linear generator of symmetry
transformations.

The form of \eq{eq:MasterSLambda} entails that an expansion of the
master equations in terms of fields leads to relatively simple
hierarchies of multi-linear equations for the expansion coefficients
of the Wilsonian effective action $S_k^{(n)}$. \Eq{eq:MasterSLambda}
is even amiable to closed partial solutions in terms of general
field dependencies.  This holds in particular for the formulation of
the mWTI as an unbroken symmetry as put forward in
\cite{Igarashi:2001mf,Sonoda:2007dj,Igarashi:2008bb,Igarashi:2009tj}.

In turn, the related master equation \cite{Wetterich:1992yh} for the
1PI effective action $\Gamma[\Phi]$, the Wetterich equation, has the
form
\begin{align}\label{eq:MasterGk}
  D[\Phi] \Gamma_k[\Phi] = \sum c_{n} \Gamma_k^{(n)}[\Phi]
  G[\Phi]\,,
\end{align}
with 
\begin{align}
G[\Phi]= \0{1}{\Gamma_k^{(2)}[\Phi]+R_k} \,,  
\label{eq:Gk}\end{align}
see also \cite{Ellwanger:1993mw,Morris:1993qb}. Note that the
propagator $G$ in \eq{eq:MasterGk} and \eq{eq:Gk} is $k$-dependent. We
drop any reference to this for the sake of a better readability of the
equations. The 1PI master equation always involves the propagator
$G[\Phi]$. The propagator relates to $S^{(2)}_k - (S^{(1)}_k)^2$ which
underlines the similarity of both sets of master equations. For more
details on progress in gauge theories including gravity we refer the
reader to the reviews
\cite{Litim:1998nf,Polonyi:2001se,Pawlowski:2005xe,Gies:2006wv,Niedermaier:2006wt,%
  Percacci:2007sz,Igarashi:2009tj,Braun:2011pp,Litim:2011cp,Rosten:2010vm,Braun:2011pp,Reuter:2012id}.

It is the inverse in \eq{eq:MasterGk} which makes solutions to
symmetry identities in a closed form less easily accessible as for the
Wilsonian effective action. Moreover, already the derivation of
symmetry identities and algebraic manipulations are structurally
simpler with \eq{eq:MasterSLambda}.

On the other hand, the numerical stability of solutions of the flow
equations of the type \eq{eq:MasterGk} is qualitatively better. Again
it is the inverse in \eq{eq:MasterGk} that triggers this
difference. For example, for large momenta the propagator $G$ decays
with the dispersion of the theory at hand. This is trivially achieved
by the form $1/(\Gamma^{(2)}_k+R_k)$ for the 1PI formulation, while it
requires non-trivial cancellations between the $S^{(2)}_k$ and
$(S^{(1)}_k)^2$ in the Wilsonian effective action framework.

The above observations on the properties of \eq{eq:MasterSLambda} and
\eq{eq:MasterGk} suggest a combined use of both equations within
aproximations to the effective actions $S_k$ and $\Gamma_k$:
\begin{itemize}
\item[(i)] One partially solves the algebraically tractable mWTI 
  based on \eq{eq:MasterSLambda}.  
\item[(ii)] The results are translated from the Wilsonian effective
  action to the 1PI effective action via the Legendre transformation
  connecting both actions. 
\item[(iii)] The flows for the correlation functions are solved in terms
  of the flow equation for the 1PI effective action.
\end{itemize} 
On the level of the correlation functions or vertices the above
strategy entails that we deduce algebraic relations between
$\Gamma_{I,k}^{(n)}\vert_{\Phi=0}$ and $S_{I,k}^{(n)}\vert_{\phi=0}$
via the Legendre transformation. This allows us to translate the
symmetry constraints for $S_{I,k}^{(n)}\vert_{\phi=0}$ to similar ones
for $\Gamma_{I,k}^{(n)}\vert_{\Phi=0}$. Finally, we use these
relations in the flow equation \eq{eq:MasterGk}. In summary this leads
to symmetry-consistent approximations to the flows of the interacting
parts of the 1PI correlation functions
$\Gamma_{I,k}^{(n)}\vert_{\Phi=0}$.

The main purpose of the present paper is to make progress on the above
programme in terms of (i) and (ii) in QED: first we provide a solution
for the mWTI for QED with a massless fermion for the Wilsonian
effective action.  This result is then translated to the 1PI effective
action. Part (iii) of the programme, the solution of the
symmetry-enhanced flow equations and its analysis will be reported in
a separate paper.

This paper is organized as follows.  In Section~\ref{sec:FunRel}, the
relation between the Wilson and 1PI effective actions as well as their
flow equations are reviewed. We also present a brief derivation of the
the modified Ward-Takahashi identity for QED.  In
Section~\ref{sec:GI+SI} we describe the truncation scheme used in the
current work.  In Section~\ref{sec:mWTI} we solve the mWTI for the
truncated Wilson action and map the result onto the 1PI effective
action. A summary and discussion are given in Section~\ref{sec:conclusion}.
Several appendices contain the technical details.

\section{Functional flows and the modified WTI}\label{sec:FunRel}

In this section we give a brief derivation of the flows for the Wilson
and one-particle irreducible (1PI) effective actions, $S_k$ and
$\Gamma_k$ respectively. These are the generating functionals of
amputated connected and 1PI correlation functions. If applied to gauge
theories, the regularisation procedure leads to modified symmetry
identities, that are also introduced here.

The derivations in the present section are kept general. In the
present work, however, we are predominantly interested in QED, so we
shall use it as an explicit example for the general relations derived
below. Applications of the FRG to Abelian gauge theories range from
QED to the Abelian Higgs models as effective theories for high $T_c$
superconductors, see e.g.\ \cite{Reuter:1994sg,Morris:1995he,%
  Bergerhoff:1995zq,Bergerhoff:1995zm,Freire:1996db,Aoki:1996fh,%
  Freire:2000sx,Kubota:2001kk,Sonoda:2007dj,Ardalan:2011ri,Fejos:2016wza}. Its
classical gauge-fixed action is given by
\begin{align}\nonumber 
  S_{\text{\tiny{cl}}}[\varphi] = &\, \014 \int_x F_{\mu\nu}^2(a)+ \int \bar\psi \left(i
    \dr(a) +i m_0\right)\psi\\[2ex]  & +\01{2\xi_0} \int_x (\partial_\mu
  a_\mu)^2-i \int_x \bar c \,\partial_\mu^2 c\,,
\label{eq:Scl}\end{align}
with 
\begin{align}\nonumber 
  \dr_\mu(a) =\,\gamma_\mu D_\mu(a)\,,\qquad D_\mu(a) = \partial_\mu + i\,
  e_0 a_\mu \,,
\end{align}
and 
\begin{align} 
  \{\gamma_\mu,\gamma_\nu\}=\,2\,\delta_{\mu\nu}\,.
\label{eq:covD}\end{align} 
The second line in \eq{eq:Scl} constitutes the gauge fixing sector
with a general covariant gauge fixing and the trivial ghost term in
QED. The classical action \eq{eq:Scl} depends on the bare couplings
(or parameters) $e_0,m_0,\xi_0$. The field $\varphi$ comprises all fields,
\begin{align}\label{eq:varphi}
  \varphi= (a_\mu,\psi,\bar\psi, c,\bar c)\,, 
\end{align}
including the decoupled ghost fields. The latter only plays a
r$\hat{\rm o}$le for the BRST transformation introduced later. 

\subsection{Wilson \& 1PI effective actions and their flows}

We consider a theory in 4-dimensional Euclidean space, which has a
gauge symmetry written as a BRST symmetry.  It is described by a
gauge-fixed action ${\cal S}[\varphi]$, a functional of fields
$\varphi^{A}$ which collectively represent gauge and matter fields as
well as ghosts and anti-ghosts.  The index $A$ denotes the Lorentz
indices of gauge fields, the spinor indices of the fermions, and other
indices distinguishing different types of generic fields.  The
Grassmann parity for $\varphi^{A}$ is expressed as
$\ep(\varphi^{A})=\ep_{A}$: $\ep_{A} =0~(1)$ if the field $\varphi^{A}$ is
Grassmann even (odd). 

In FRG approach, we introduce an IR cutoff $k$ through positive
functions that behave as
\begin{align}
 \quad K^{A}\Bigl(\frac{p}{k}\Bigr)\rightarrow \quad  \left\{
		\begin{array}{ll}
		 1 & (p^2 < k^2) \\[2ex]
		  0 & (p^2 > k^2 )~.
		\end{array}
               \right.
\label{cutoff-func}
\end{align}
The functions go to zero sufficiently rapidly as $p^{2} \rightarrow
\infty$.  For simplicity, we write the functions as $K^{A}(p)$ in the
rest of the paper. The action ${\cal S}$ defined at some UV scale is
given as a sum the kinetic and interaction terms,
\begin{eqnarray}
{\cal S}[\varphi] = \frac{1}{2}\varphi \cdot D \cdot \varphi + 
{\cal S}_{I}[\varphi]\,, 
\label{micro-action}
\end{eqnarray}
where we have used a condensed matrix notation in momentum space.  For
example, the kinetic term in \eq{micro-action} has the explicit
form
\begin{align}
  \varphi \cdot  D  \cdot \varphi = \int \frac{d^{4}p}{(2\pi)^4} 
  \varphi^{A}(-p) D_{AB}(p) \varphi^{B}(p)\,.
\label{cond-not}
\end{align}
Due to the presence of the regulator functions \eq{cutoff-func} the
Wilsonian effective action $S_k$ is the scale-dependent
generating functional of amputated connected correlation functions.
The scale-dependent generating functional for one-particle irreducible
(1PI) correlation functions, $\Gamma_k$ is obtained via a
Legendre transform from $S_k$. The scale-dependence of
$S_k,\Gamma_k$ is encoded in flow equations, i.e., the
Polchinski equation and Wetterich equation respectively.

In the main text of the present work we resort to normalised fields
and couplings in the Wilson effective action $S_k$,  
\begin{align}
\bar\phi &= Z^{1/2}_\phi \phi\,,\nn\\
g &= Z_{g}\, g_{0}\,,
\label{eq:barphi}\end{align}
where $g_{0}$ are $k$-independent couplings defined at some UV scale.
For QED with the classical action \eq{eq:Scl} we have the parameters
$\vec g_0=(e_0,m_0,\xi_0)$ and the normalised couplings read
\begin{align}
e &= Z_{e}\, e_{0}\,,\qquad m=Z_m m_0\,,\qquad \xi=Z_\xi \xi_0\,. 
 \label{eq:gg0QED}
\end{align}

The rescaling \eq{eq:barphi} facilitates the access to scaling
properties, as the scale-dependent effective actions $S_k$ and
$\Gamma_k$ have the same renormalisation group equation as the full
effective actions at $k=0$, see \cite{Pawlowski:2005xe}, with
\begin{align}\label{eq:muRG}
\mu \0{d }{d\mu} \,\rm GF=0\,,\quad {\rm where} \quad 
{\rm GF } = S_k,\Gamma_k, S_0,\Gamma_0\,.
\end{align}
Moreover, the related expansion coefficients in powers of $\bar\Phi$,
the amputated connected correlation functions or their 1PI parts, are
renormalisation group invariant.

For the sake of accessibility of the current work and due to the minor
modifications of the derivation in the presence of the rescaled fields
we recall the derivation of the flow of the Wilson effective action in
Appendix~A. Since the Wilson action is a functional of $(Z^{A})^{1/2}
\phi^{A}$, we may extract contributions of the anomalous dimensions from
$S_{k}(Z^{1/2}\phi)$.  The final equation for $S_k$ in terms of the
rescaled fields $\bar\phi$ reads, see \eq{PFE-total-Z fixedapp},
\begin{widetext}
\begin{align}
\pa_{t} S_{k}\mid_{\bar\phi} 
= - \bar \phi^{A} \Bigl[\left(\pa_{t} \log K\right)^{A} - \012 \eta_{A}\Bigr]
\frac{\pa^{l} S_{k}}{\pa\bar \phi^{A}}
+ (-)^{\ep_{A}}
\frac{1}{2} 
\biggl[\Bigl\{\pa_{t}K -\eta K(1-K)\Bigr\}D^{-1}
\biggr]^{AB}
\biggl[
\frac{\pa^{l} S_{k}}{\pa\bar\phi^{B}}\frac{\pa^{r} S_{k}}{\pa\bar\phi^{A}}
- \frac{\pa^{l} \pa^{r} S_{k}}{\pa\bar\phi^{B} \pa\bar\phi^{A}} 
\biggr]\,,
\label{PFE-total-Z fixed} \end{align}
\end{widetext}
with the anomalous dimensions 
\begin{eqnarray}
\eta_{A} = - \pa_{t} \log Z^{A}\,,
\label{anomalous-dim}
\end{eqnarray}
for $\phi^{A}$.  For the flow equation of the interaction part
$S_{I,k}$, see (\ref{PFE-intapp}) in the Appendix~A.

The Wilson action $S_{k}$ is the generating functional of the connected
(amputated) cutoff Green functions. Already its one-particle irreducible
(1PI) part, the scale-dependent effective action, $\Gamma_{k}$, carries
all the information about the correlation functions of the theory at
hand. It is a part of the full Legendre effective action, $\tilde \Gamma_k$,
which is obtained via the Legendre transformation of the Wilson
effective action:
\begin{eqnarray}
  \tilde \Gamma_{k}[\bar\Phi] &=& \Gamma_{k}[\bar \Phi]
  + \frac{1}{2} \bar \Phi \cdot R_k \cdot \bar \Phi\,.
\label{Legendre1}
\end{eqnarray}
It is the standard scale-dependent effective action, $\Gamma_k$, that
is used in most applications.  The regulator $R_k$ relates to $K$ with
\begin{eqnarray}
R_{k} = \frac{K}{1-K} D\,. 
\label{definition of R}
\end{eqnarray}
A particularly concise iterative relation exists between the
interaction part of $\Gamma_{k}$,
\begin{align}\label{GI}
  \Gamma_{I,k}[\bar\Phi]=\Gamma_{k}[\bar\Phi]-\012 \bar \Phi \cdot D
  \cdot \bar\Phi\,,
\end{align} 
and the interaction part of $S_{k}$, 
\begin{align}
 \Gamma_{I,k}[\bar \Phi] = S_{I,k}[\bar\phi]
  - \frac{1}{2}  
(\bar \Phi - \bar \phi) \cdot \bigl(G^{(0)}\bigr)^{-1}
\cdot (\bar \Phi - \bar \phi) \,,
\label{Legendre2}
\end{align}
where $\bigl(G^{(0)}\bigr)^{AB} = \bigl((1-K)D^{-1}\bigr)^{AB}$ are
the high momentum propagators. This relation is most conveniently
written in terms of the two-point functions of the Wilson effective
action and the 1PI effective action. They comprise the propagators of
the theory at hand. These two-point functions are the basic building
blocks of the functional flow equations and read
\begin{align}
  \left(S_{I,k}^{(2)}[{\bar \phi}] \right)_{AB} \equiv \frac{\pa^{l}
    \pa^{r} S_{I,k}}{\pa {\bar \phi}^{A}\pa {\bar \phi}^{B}}\,,\quad
  \left(\Gamma_{k}^{(2)}[{\bar \Phi}] \right)_{AB} \equiv
  \frac{\pa^{l} \pa^{r} \Gamma_{k}}{\pa {\bar \Phi}^{A}\pa {\bar
      \Phi}^{B}} \,,
\end{align}
where $\partial^l$ and $\partial^r$ stand for left- and right-derivatives
respectively.  Then, \eq{Legendre2} leads us to the relation
\begin{align}\label{relation of 2nd derivatives}
S_{I,k}^{(2)}[{\bar \phi}]
= (G^{(0)})^{-1}-(G^{(0)})^{-1}\cdot G[\bar\Phi] \cdot (G^{(0)})^{-1} \,, 
\end{align} 
with the full propagator 
\begin{align}
G[\bar\Phi] = \bigl({\Gamma_{k}^{(2)}[{\bar \Phi}]+R_k} \bigr)^{-1}, 
\label{full propagator}
\end{align}
see also \eq{eq:Gk}. Note that we recover $G^{(0)}$ if we drop the
contributions from the interaction part in \eq{full propagator}.
Using \eq{relation of 2nd derivatives} and the relation between
$\bar\Phi^{A}$ and $\bar\phi^{A}$,
\begin{eqnarray}
  \bar  \Phi^{A} =
  \bar \phi^{A} 
- \bigl(G^{(0)}\bigr)^{AB}
\frac{\pa^{l} S_{I,k}}{\pa \bar \phi^{B}}\,,
\label{field-relation}
\end{eqnarray}
the flow equation for $S_k$ is easily transformed into one for
$\Gamma_k$,
\begin{align}\label{eq:flowG}
  \left( \partial_t -\012 \eta_{A}\, \bar\Phi_A
    \0{\delta}{\delta\bar\Phi_A}\right) \Gamma_k[\bar\Phi] = \012 {\rm
    Tr}\, G[\bar\Phi]\cdot \left( \partial_t- \eta\right)\cdot R_k
  \,.
\end{align}
\Eq{eq:flowG} shows the relation between derivatives of $\Gamma_k$ and
the inverse of the second field derivative of ${\tilde \Gamma}_k$
already discussed in the introduction. This structure is also present
in the mWTI, and suggests a solution of the mWTI in terms of $S_k$,
and its insertion into $\Gamma_k$.

Finally we remark that the Wilson action $S_{I,k}$ can be
iteratively expanded in terms of the Legendre action
$\Gamma_{I,k}$, their field derivatives and the cutoff (high
energy) propagators $G^{(0)}$:~
\begin{eqnarray}\nonumber 
  S_{I,k}[\bar\phi] &=& \Gamma_{I,k}\left[ \bar\phi- 
    G^{(0)}\cdot S_{I,k}^{(1)}\right] + \frac{1}{2}  S_{I,k}^{(1)} \cdot 
  G^{(0)}\cdot S_{I,k}^{(1)}\\[2ex] \nonumber 
  & =&  \Gamma_{I,k}[\bar\phi] - \frac{1}{2}
  \Gamma_{I,k}^{(1)}\cdot G^{(0)}\cdot\Gamma_{I,k}^{(1)}\\[2ex] 
  & & 
  + \frac{1}{2} \Gamma_{I,k}^{(1)}\cdot G^{(0)} \cdot \Gamma_{I,k}^{(2)} \cdot 
  G^{(0)}\cdot \Gamma_{I,k}^{(1)}  
  + \cdots\,.
\label{tree-ex}
\end{eqnarray}
The tree expansion \cite{Ishikake:2005rk} in \eq{tree-ex} will be used
to construct $S_{I,k}$ from $\Gamma_{I,k}$ in QED.

\subsection{Derivation of the mWTI}

From now on we concentrate on QED. In the present section we briefly
recapitulate the derivation of the mWTI. The BRST transformations
$\delta\varphi^{A}$ take the form,
\begin{eqnarray}
  \delta\varphi^{A} = (R_1)^{A}_{~B}\varphi^{B} + (R_2)^{A}_{~B}\varphi^{B}c~,
\label{linear-BRS}
\end{eqnarray}
where $(R_1)^{A}_{~B}$ and $(R_2)^{A}_{~B}$ are field independent
functions and $c$ is the ghost field.  The classical
BRST transformations of the photon and the fermion fields are
described by the first and the second terms in \eq{linear-BRS},
respectively.  Note that the transformations in \eq{linear-BRS} are
linear in the field except for the presence of a free ghost field
denoted as $c$.

Even for linear gauge symmetries, the BRST transformations for the IR
fields $\delta\phi^{A}$ become non-linear: the BRST transformations
for the IR fields get modified due to interactions generated by the
integration over the higher momentum modes. The free ghost and the
anti-ghost field do not contribute to the modifications. The full
BRST transformation of the IR field $\bar \phi$ can be rewritten similarly
to the classical transformation \eq{linear-BRS} in terms of the 
mean field $\bar \Phi$ in \eq{field-relation}: 
\begin{align}
  \delta\bar \phi^{A} = K (R'_1)^{A}_{~B} \bar \Phi^{B}  + K
  (R'_2)^{A}_{~B} \bar \Phi^{B} c\,.
\label{delta-phi-composite0}
\end{align}
This has been detailed in \cite{Igarashi:2009tj} in terms of a composite
field language; see Appendix~\ref{app:WTI} for a
few details, including the changes of the coefficients from
\eq{linear-BRS} to \eq{delta-phi-composite0} owing to the wave
functions renormalisation.  Using \eq{IR-WT-op} and
\eq{delta-phi-composite0}, we obtain the mWTI
\begin{align} 
\Sigma_{k}[\bar \phi] = S^{(1)}_k\cdot \delta
  \bar \phi + \Tr \, K\cdot  R'_2 \cdot G^{(0)}_k \cdot
    S^{(2)}_k\cdot c=0\,.
\label{Sigma}
\end{align}

\subsection{The mWTI for the Wilson action of QED}

So far we have briefly recalled the mWTI and its derivation. Now we
turn to their application to QED in a given approximation. To this
end, we first fix the coefficient functions $R^{A}_{~B}$ and
$R^{A}_{~BC}$ from the classical BRST transformations for the UV
fields $\delta \varphi^{A} = R^{A}_{~B}c^{B} + R^{A}_{~BC} \varphi^{B}
c^{C}$:
\begin{eqnarray}
\delta_{cl}~ a_{\mu}(p) &=& -i p_{\mu} c(p)\,,\nn\\[2ex]
%\delta_{cl}~ b(p) &=& 0\,,\nn\\[2ex]
\delta_{cl}~ \psi(p) &=& -i~e_{0}~\int_{q} \psi(q) c(p-q)\,,\nn\\[2ex]
\delta_{cl}~ {\bar \psi}(-p) &=& i~e_{0} 
\int_{q}  {\bar \psi}(-q) c(q-p)\,,\nn\\[2ex]
 \delta_{cl}~ {\bar c}(p) &=&
%  - b(p)
 \xi_{0}^{-1} p_{\mu}~a_{\mu}(p) \,,
\label{classical-BRS}
\end{eqnarray}
where the bare gauge coupling $e_{0}$ and gauge parameter $\xi_{0}$ are
$k$ independent constants.  Then as shown in Appendix B2, the quantum BRST
transformations derived from \eq{delta-phi-composite0} are given for
the renormalised fields
\begin{align}\label{eq:barfields} \bar\phi =
(a,\psi,\bar\psi, c,\bar c)\,.
\end{align} 
The ghost and anti-ghost are free fields, and have no genuine wave
function renormalisation. This entails the natural choice $Z_c=Z_{\bar
  c}=1$. For convenience we choose
\begin{align}\label{eq:Zc}
Z_c= 1/Z_e\,,\quad \quad Z_{\bar c}=Z_e\,,\quad {\rm with} \quad e = Z_e\, e_0\,,
\end{align} 
With this rescaling we are lead to
\begin{eqnarray}
\delta a_{\mu}(p) &=& - K(p)~Z_e Z_3^{1/2} \,i p_{\mu} C(p)~,\nn\\[2ex]
\delta \psi(p) &=& -K(p)~i~e  
\int_{q} \Psi(q) C(p-q)~,\nn\\[2ex]
\delta  {\bar \psi}(-p) &=& K(p)~i~e 
\int_{q}  \bar \Psi(-q)  
C(q-p)\,, \nn \\[2ex] 
\delta  {\bar c}(p) &=& K(p)~
Z_e Z_{3}^{1/2}\,\xi^{-1} p_{\mu}~A_{\mu}(p)\,, 
\label{quantum-BRS}
\end{eqnarray}
in terms of the composite mean fields, 
\begin{align}\label{eq:barmf} 
\bar \Phi=(A_\mu,
\Psi,\bar\Psi,C,\bar C)\,,
\end{align} 
defined in \eq{field-relation}. In \eq{quantum-BRS} the wave function
renormalisations are absorbed in the fields. The only remnant is 
the product $Z_e\,Z_3^{1/2}$ of the coupling renormalisation
function and the wave function renormalisation of the photon. If this
product is set to unity we are left with the classical transformation
except for the occurrence of the composite mean fields on the right
hand side. Hence, the rescaling \eq{eq:Zc} make already apparent the
standard relations. The composite mean fields in QED follow from their
general definition in \eq{field-relation} as
\begin{eqnarray} \nonumber 
 A_{\mu}(p)&=& K^{-1}(p)~a_{\mu}(p)
  - (G_{G}^{(0)})_{\mu\nu}
\frac{\pa S_{k}}{\pa a_{\nu}(-p)}
\,,
\end{eqnarray} 
\begin{eqnarray}\nonumber 
\Psi(p) &=& 
 K^{-1}(p) \psi(p) 
- G_{F}^{(0)}\frac{\pa^{l} S_{k}}{\pa {\bar\psi}(-p)}\,,
\\
\bar\Psi(-p) &=& 
K^{-1}(p) \bar\psi(-p) - 
\frac{\pa^{r} S_{k}}{\pa \psi(p)}
G_{F}^{(0)} ~.
\label{eq:QED-composite}
\end{eqnarray}
Note that $C(p) =c(p)$, $\bar C(p) = \bar c(p)$ with the same wave
function renormalisation \eq{eq:Zc}, since $c$ and $\bar c$ are free
fields. From \eq{Sigma}, we obtain the WT operator for QED,
\begin{widetext}
\begin{align} \nn
  \Sigma_{k}[\bar \phi] = &\, Z_e Z_3^{1/2} \int_{p}\biggl[\frac{\pa
    S_{k}}{\pa a_{\mu}(p)}(-ip_{\mu})c(p) + \frac{\pa^{r} S_{k}}{\pa
    {\bar c}(p)} \01\xi p_{\mu} a_{\mu}(p) \biggr]\\[2ex] 
&\, -i~e \int_{p,q}
  \biggl[\frac{\pa^{r} S_{k}}{\pa \psi_{\al}(q)}\frac{K(q)}{K(p)}
  \psi_{\al}(p) -\frac{K(p)}{K(q)}{\bar \psi}_{\hat\al}(-q)
  \frac{\pa^{l} S_{k}}{\pa {\bar \psi}_{\hat\al}(-p)} \biggr]c(q-p)\nn\\[2ex]
  & -i~e \int_{p,q} U_{\beta\hat\al}(-q,p) \biggl[ \frac{\pa^{l}
    S_{k}}{\pa {\bar \psi}_{\hat\al}(-p)} \frac{\pa^{r} S_{k}}{\pa
    \psi_{\beta}(q)} - \frac{\pa^{l}\pa^{r} S_{k}}{\pa {\bar
      \psi}_{\hat\al}(-p)\pa \psi_{\beta}(q)} \biggr]c(q-p)~,
 \label{WT-QED1}
\end{align}
\end{widetext}
where
\begin{align}
 U(-q,p)_{\beta\hat\al}
=&\, \Bigl[K(q)G_{F}^{(0)}(p) - K(p) 
G_{F}^{(0)}(q)\Bigr]_{\beta\hat\al} \, .
\label{U matrix 2}
\end{align}
The bare gauge parameter $\xi_{0}= \xi/Z_3$ and the gauge coupling
$e_0 = e/Z_e$ are $k$-independent and one can show that
$\Sigma_{k}[\phi]$ in \eq{WT-QED1} itself is a composite operator as
defined in \cite{Igarashi:2009tj,Pawlowski:2005xe}, and satisfies the
flow equation for composite operators discussed in
Appendix~\ref{app:composite}.

\section{The truncated Wilson action}\label{sec:GI+SI}
In this section, we construct the interaction part of the Wilson
action, $S_{I,k}$, within a suitable truncation. Since $S_{I,k}$ is
related to the generating functional of connected Green's function
obtained by integrating over high energy modes of the original theory,
it can be constructed in terms of its 1PI part, $\Gamma_{I,k}$.
The latter is quite useful for the study of the RG flow.  Therefore,
we first provide a truncated form of $\Gamma_{I,k}$ and then
construct $S_{I,k}$ via the Legendre transformation or equivalently
the tree expansion (\ref{tree-ex}).

In addition to corrections to the primitively divergent correlation
functions, the two-point functions and the gauge coupling, we also
introduce the four-fermi couplings. These couplings are the lowest
order of higher dimensional matter interactions, that are generated 
from the primitively divergent correlation functions within one RG-step. Here we
need them in order to close our approximation scheme. In terms of the
interaction part $\Gamma_{I,k}$ of the 1PI effective action this
truncation corresponds to corrections to the electron and photon
two-point functions, the self-energy, $h^{(\bar\psi\psi)}$, and the vacuum
polarisation, $h^{(aa)}$, respectively, the electron-photon vertex and
its quantum corrections, $h^{(\bar\psi a\psi)}$, and the four-electron
scattering vertex $h^{(\bar\psi\psi\bar\psi\psi)}$. Schematically this
leads to
 \begin{align} 
   \nn \Gamma_{I,k}[\bar\Phi] = &\,\012 h^{(aa)} \cdot A^2 +
   h^{(\bar\psi\psi)} \cdot \bar\Psi \Psi \\[2ex]  & + h^{(\bar\psi a
     \psi)} \cdot \bar\Psi A\Psi + h^{(\bar\psi\psi\bar\psi\psi)}
   \cdot \bar\Psi \Psi \bar\Psi\Psi\,,
\label{eq:GIschem}\end{align}
where the powers in the fields in \eq{eq:GIschem} stand for the tensor
products. The first line comprises the corrections to the kinetic
terms of photon and electrons, and the second term comprises the
interaction terms. The full expression including all momentum
dependencies and Lorentz indices is given in
Appendix~\ref{app:Wilson}. Correspondingly, the schematic expression
for the interaction part of the Wilson effective action reads
 \begin{align} 
   \nn S_{I,k}[\bar \phi] =& \012 ( \bar G_G^{-1} \cdot \tilde
   h^{(aa)}) \cdot (\bar G_G a\,\, \bar G_G a) \\[2ex] \nn & + (\bar
   G_F^{-1}\cdot \tilde h^{(\bar\psi\psi)}) \cdot (\bar G_F
   \bar\psi\,\, \bar G_F \psi )\\[2ex] \nn & + \tilde h^{(\bar\psi a
     \psi)}\cdot ( \bar G_G a \,\, \bar G_F\bar\psi\,\, \bar
   G_F\,\psi)\\[2ex]  &+ \tilde h^{(\bar\psi\psi\bar\psi\psi)} \cdot (\bar
   G_F\bar\psi \,\, \bar G_F\psi\, \, \bar G_F\bar\psi \,\, \bar
   G_F\psi)\,,
\label{eq:SIschem}\end{align}
and the explicit expression with momentum dependence and Lorentz
indices can be found in Appendix~\ref{app:Wilson}, \eq{eq:SIk}.  The
form \eq{eq:SIschem} makes apparent that the Wilson effective
action generates amputated connected correlation functions. The
explicit $\bar G$-factors are arranged such that the 1PI-parts of the
vertex coefficient functions $\tilde h$ agree with the 1PI coefficient
functions $h$ in \eq{eq:GIschem}. Hence, for connected correlation
functions that only contain 1PI parts, $h$ and $\tilde h$ agree. This
holds true for two and three-point functions, see also the relations
\eq{h-H relations}. In turn, for higher correlation functions, $\tilde
h$ can be easily expanded in $h$ according to the diagrammatic
expressions of connected correlation functions in terms of 1PI
correlations. Due to the definition of the fields the propagators in
\eq{eq:SIschem} are normalised with the inverse bare propagators,
leading to
\begin{eqnarray}
{\bar G}(p) \equiv 
G(p)
\bigl( G^{(0)}(p) \bigr)^{-1} \,, 
\label{ratio}
\end{eqnarray}
where the $k$-dependence of $\bar G$ is implicit.  For photons and
electrons this reads more explicitly
\begin{eqnarray}
\bigl(\bar{G}_{G}\bigr)_{\mu\nu}(p) &=& 
\biggl[\frac{1}{{\bf 1} + h^{(aa)} G_{G}^{(0)}}\biggr]_{\mu\nu}(p) \,,
\nn \\[2ex] 
\bigl(\bar{G}_{F}\bigr)_{\alpha\beta}(p) &
= &\biggl[\frac{1}{{\bf 1} + G_{F}^{(0)}h^{(\bar\psi\psi)}}\biggr]
_{\alpha\beta}(p) \,.
\label{eq:barpinvrop}
\end{eqnarray}
As discussed above we have for the two point functions
\begin{subequations} 
\begin{align}
& \tilde{h}^{(aa)}_{\mu\nu}(p) = {h}^{(aa)}_{\mu\nu}(p) \,,\qquad 
\tilde{h}^{(\bar\psi\psi)}_{\hat\alpha\alpha}(p) = 
{h}^{(\bar\psi\psi)}_{\hat\alpha\alpha}(p) \,,
\label{hH2}\end{align} 
and for the three point function
\begin{align}
  \tilde{h}^{(\bar\psi a\psi)}_{\hat\alpha\alpha,\mu}(p_{1},p_{2})=
  {h}^{(\bar\psi a\psi)}_{\hat\alpha\alpha,\mu}(p_{1},p_{2}) =
  {h}_e(p_{1},p_{2}) \gamma_{\mu,\hat \alpha\alpha} \, .
\label{hH3}\end{align} 
For the four-fermi interaction we have a 
genuine connected part which is not 1PI:
\begin{align}
&\tilde{h}^{(\bar\psi\psi\bar\psi\psi)}_{\hat\alpha\alpha\hat\beta\beta}
(p_{1},p_{2},p_{3}) = {h}^{(\bar\psi\psi\bar\psi\psi)}
_{\hat\alpha\alpha\hat\beta\beta}(p_{1},p_{2},p_{3})\nn \\[2ex] 
 &  - \frac{1}{2}
h^{(\bar\psi a\psi)}_{\hat\alpha\alpha,\mu}(p_{1},p_{2})  
h^{(\bar\psi a\psi)}_{\hat\beta\beta,\nu}(p_{3},p_{4})
\bigl(G_{G}\bigr)_{\mu\nu}(p_1+p_2)\,.
\label{hH4}\end{align}
\label{h-H relations}\end{subequations} 
The 4-fermi term ${\tilde h}^{(\bar\psi\psi\bar\psi\psi)}$ contains a
one-photon exchange contribution with two 3-point vertices as given in
the last term in \eq{h-H relations}, in addition to the one
proportional to 4-point function $h^{(\bar\psi\psi\bar\psi\psi)}$.

In this paper we concentrate on massless fermions, and hence we have
chiral symmetry. Furthermore, we only take into account the classical
tensor structure of the photon-electron vertex, $\gamma_\mu$, i.e.,
\begin{align}
 & {h}^{(\bar\psi\psi)}_{\hat\alpha\alpha}(p) = \sigma(p)\Slash{p}, \nn\\[2ex]
 & {h}^{(\bar\psi a\psi)}_{\hat\alpha\alpha,\mu}(p_{1},p_{2}) =
     {h}_e(p_{1},p_{2}) \gamma_{\mu,\hat \alpha\alpha} \, ,
\label{h-approximation}
\end{align}
where $\sigma(p)$ and $h_e(p_1,p_2)$ are form factors. In \eq{hH4}, we
have not specified the form of the four-fermi interaction. In the
present work we take the standard chiral form which already follows
from one RG-step. It can be written as a combination of two terms.
The first one reads
\begin{align} 
& \frac{1}{2k^{2}} \int_{p_{1},p_2,p_{3}}\biggl[ 
h_{S}(s,t,u)
\Bigl\{\left(
{\bar\Psi}(p_{1})\Psi(p_{2})\right)\left({\bar\Psi}(p_{3})\Psi(p_{4})\right) \nn\\[2ex]
&\hspace{3cm}
- \left({\bar\Psi}(p_{1})\gamma_{5}\Psi(p_{2})\right)\left({\bar\Psi}(p_{3})
\gamma_{5}\Psi(p_{4})\right)\Bigr\}
\nn\\[2ex] 
& 
\hspace{1cm}+ h_{V}(s,t,u)\Bigl\{\left(
{\bar\Psi}(p_{1})\gamma_{\mu}\Psi(p_{2})\right)
\left({\bar\Psi}(p_{3})\gamma_{\mu}
\Psi(p_{4})\right)
\nn\\[2ex] 
&\hspace{1.5cm}
+ \left({\bar\Psi}(p_{1})\gamma_{5}\gamma_{\mu}\Psi(p_{2})\right)
\left({\bar\Psi}(p_{3})
\gamma_{5}\gamma_{\mu}\Psi(p_{4})\right)\Bigr\}
\biggr]\,,
\label{eq:4fermi1}
\end{align} 
while the second one follows as 
\begin{align}
\nn
& \frac{1}{2k^{4}}\int_{p_{1},p_2,p_{3}} 
h_{V'}(s,t,u)(p_{1}+p_{4})_{\mu}(p_{2}+p_{3})_{\nu}\nn\\[2ex] 
& \hspace{2cm} \times\biggl[ 
\left({\bar\Psi}(p_{1})\gamma_{\mu}\Psi(p_{2})\right)
\left({\bar\Psi}(p_{3})\gamma_{\nu}
\Psi(p_{4})\right) 
\nn\\[2ex] 
&
\hspace{2cm}+ \left({\bar\Psi}(p_{1})\gamma_{5}\gamma_{\mu}\Psi(p_{2})
\right)
\left({\bar\Psi}(p_{3})\gamma_{5}\gamma_{\nu}
\Psi(p_{4})\right)
\biggr]~.
\label{Gamma_I}
\end{align}
Here we have included the chiral invariant 4-fermi interactions with
form factors which are functions of $s=(p_{1}+p_{2})^{2},
t=(p_{1}+p_{3})^{2}$, and $u=(p_{1}+p_{4})^{2}$.  As 4-fermi terms, we
introduced a derivative vector coupling other than commonly used
scalar and vector couplings.  We will see shortly how these higher
dimensional operators with the form factors affect relations among
lower dimensional operators via non-trivial loop contributions in the
mWTI. From \eq{Gamma_I} with the use of the Legendre transformation,
we obtain the interaction part of the truncated Wilson action which is
given in Appendix~\ref{app:Wilson}.

\section{Constraints from the modified WTI}\label{sec:mWTI}

Having constructed the Wilson action \eq{SI}, we now derive the
relations for the couplings, resulting from the mWTI, $\Sigma_{k}
=0$. When the WT operator $\Sigma_{k}$ is expanded as polynomials of
the fields, $\Sigma_{k} =0$ leads to a number of coupled relations for
the couplings and form factors in the Wilson action. We substitute
$S_{I,k}$ given by \eq{SI} into \eq{WT-QED1} and find coefficients of
operators $a_\mu c$ and ${\bar \psi}\psi c$ in $\Sigma_k=0$.  In this
manner, we obtain two WT relations \eq{2nd relation1} and \eq{1st
  relation1}. Furthermore, we assume locality of the fermionic
bilinear term and the gauge interaction
\begin{eqnarray}
\sigma(p) =0, \quad h_{e}(p,q) =1.
\label{assumption}
\end{eqnarray}
The first WT relation out of $a_\mu c$ terms is given by
\begin{align} 
p_{\mu}h^{(aa)}_{\mu\nu}(p)=
 p_{\nu}{\cal L}(p) 
= 
%e_{0} e
 \0{e^{2}}{Z_{e} Z_{3}^{1/2}}
\int_{q}~{\rm Tr}\left[U(p-q,q)\gamma_{\nu}\right]
,
\label{1st relation}
\end{align}
where ${\cal L}$ is the longitudinal part of $h^{(aa)}_{\mu \nu}$
\begin{eqnarray}
h^{(aa)}_{\mu\nu}(p) = P^{T}_{\mu\nu}{\cal T}(p) + P^{L}_{\mu\nu}{\cal L}(p)~.
\label{decomposition}
\end{eqnarray}

We next consider $\bar\psi \psi c$ terms and obtain the second WT
relation as shown in Appendix C,
\begin{widetext}
\begin{eqnarray}
&& \left(e - e Z_{e} Z_{3}^{1/2} \right)({\Slash{p}}- {\Slash{q}})
 -   2e \int_{l}\biggl[ \frac{1}{k^{2}}\Bigl\{
\bigl(h_{S} - 2h_{V}\bigr)\bigl(l^{2},(p+q+l)^{2},(p-q)^{2}\bigr)\nn\\[2ex] 
&&-2h_{V}\bigl((p-q)^{2},(p+q+l)^{2},l^{2}\bigr)\Bigr\}
+ e^{2} \bigl(1-K(l)\bigr) T(l^{2}) 
+ \frac{1}{k^{4}} \Bigl\{2(p-q)^{2}\nn\\[2ex] 
&& \times h_{V'}\bigl(l^{2},(p+q+l)^{2},(p-q)^{2}\bigr)
+ l^{2}h_{V'}\bigl((p-q)^{2},(p+q+l)^{2},l^{2}\bigr)
\Bigr\}\biggr]U(-q-l,p+l)
 \nn\\[2ex] 
&& -e  \int_{l}\biggl[\frac{2}{k^{4}}
h_{V'}\bigl((p-q)^{2},(p+q+l)^{2},l^{2}\bigr) +
 e^{2} \frac{(1-K(l))}{l^{2}}
\{T(l^{2}) - {\xi} L(l^{2})\}
\biggr]\nn\\[2ex] 
&& \times {\Slash{l}}U(-q-l,p+l){\Slash{l}}
=0~.
\label{2nd relation2}
\end{eqnarray}
\end{widetext}
%%%%%%%%%%%%%%
In writing \eq{2nd relation2}, we used the expressions for the
gauge full propagator as well as $\left({\bar G}_G\right)_{\mu\nu}$
defined in \eq{ratio},
\begin{eqnarray} \nn 
&& (G_G)_{\mu\nu}(p) =  
\bigl(1-K(p)\bigr) 
\Bigl(
T(p) P^{T}_{\mu\nu}
+ \xi L(p) P^{L}_{\mu\nu}\Bigr)~,
\\[2ex]
&&(\bar G_G)_{\mu \nu}(p) =
p^{2}\biggl(T(p)P^{T}_{\mu\nu} + L(p)P^{L}_{\mu\nu}\biggr)~,
\label{decom}
\end{eqnarray} 
where 
\begin{align}
T(p) = \frac{1}{ p^{2} +(1-K){\cal T}}\,, &\qquad 
L(p) = \frac{1}{p^{2} +\xi (1-K){\cal L}}\,.
\label{T and L}
\end{align}
%%%%%%%%%%%%%
Note that \eq{2nd relation2} consists of terms with and without
the integration over the momentum $l$.  The first term without the
integration is the tree term and the rest is one-loop terms.  It is
important to realize that we call a term as the tree or loop term in
reference to the Wilson action $S_k$ that is obtained after
integrating out all the modes with their momenta above the cutoff.
Even a tree term contains the contributions of the higher momentum
modes.

The first term in \eq{2nd relation2} is proportional to
$(\Slash{p} - \Slash{q})$ with a momentum independent factor. In
contrast to this tree term, the rest consist of loop integrals which
lead to some functions of momenta $p$ and $q$.  It is reasonable
therefore to require that these two different kinds of contributions
vanish separately.

Vanishing of the first term
gives 
\begin{eqnarray} 
Z_{e}~Z_{3}^{1/2} = 1~,
\label{Ze=sqrtZ3}
\end{eqnarray}
where the constant $Z_{e}$ for finite renormalisation of the gauge coupling 
may be defined as $e = e(k) = Z_{e} e(\mu) = Z_{e} e_{0}$. 
This corresponds exactly to the well-known identity
\begin{align} 
 {Z}_{1} \equiv {Z}_{2}{Z}_{3}^{1/2} {Z}_{e}
  = {Z}_{2}~. 
\label{Z1=Z2'}
\end{align}
Therefore, the standard relation ${Z}_{1}={Z}_{2}$ which ensures the
charge universality remains unchanged in our realisation of gauge
symmetry in QED.

On the other hand, there are two independent integrals containing $U$
and $\Slash{l}U\Slash{l}$. If we demand the integrands to vanish we
arrive at two non-trivial constraints:
\begin{widetext} 
\begin{eqnarray}
&& \frac{1}{k^{2}}\biggl\{
\bigl(h_{S} - 2h_{V}\bigr)\bigl(l^{2},(p+q+l)^{2},(p-q)^{2}\bigr)
-2h_{V}\bigl((p-q)^{2},(p+q+l)^{2},l^{2}\bigr)
\biggr\}\nn\\[2ex] 
&& \hspace{3cm}  = e^{2} \bigl(1-K(p-q)\bigr)
\Bigl\{T\bigl((p-q)^{2}\bigr) - \xi L\bigl((p-q)^{2}\bigr)
\Bigr\}
-\frac{e^{2}}{2}
\Bigl\{{T}({l}^{2}) + \xi {L}({l}^{2})\Bigr\}\,,
\end{eqnarray}
\end{widetext} 
and 
\begin{eqnarray} \nn 
 && \frac{1}{k^{4}}h_{V'}\bigl((p-q)^{2},(p+q+l)^{2},l^{2}\bigr)\\[2ex] 
 & & \hspace{2cm}= - e^2 \frac{2 l^2}{1-K(l)}
\Bigl\{{T}({l}^{2}) - \xi {L}({l}^{2})\Bigr\}\,,
\label{mom.depend.relation}
\end{eqnarray}
where \eq{Ze=sqrtZ3}, a result from the 2nd WT relation, is used.
These imply that the form factors $h_{S}$ and $h_{V}$ are functions of
two variables, $h_{S}(s,t,u) = h_{S}(s,u)$ and $h_{V}(s,t,u)
=h_{V}(s,u)$, while $h_{V'}$ is a function of a single variable,
$h_{V'}(s,t,u)=h_{V'}(u)$. It is now clear that the momentum dependent
form factors in the 4-fermi interactions are needed to cancel the
photon exchange contributions in the mWTI for the present assumptions
$\sigma =0$ and $h_{e}=1$ in \eq{assumption}.

Let us turn to the first WT relation (\ref{1st relation}). For a
specific choice of the cutoff function
\begin{eqnarray}
K(p) = \exp(-p^{2}/k^{2})~,
\label{gaussian}
\end{eqnarray}
the r.h.s. of \eq{1st relation} can be calculated analytically
as shown in Appendix~\ref{app:der2}. We obtain the analytic expression of the
longitudinal component of the photon 2-point function
\begin{eqnarray}\nn 
{\cal L}(p)& =&
- e^{2} \frac{k^{2}}{2\pi^{2}{\bar p}^{4}}
\biggl[ 1 - \exp(-{\bar p}^{2}/2) -\\[2ex] 
&& \hspace{2cm} {\bar p}^{2}
\Bigl(1 - \frac{1}{2}\exp(-{\bar p}^{2}/2) \Bigr)
\biggr]~, 
\label{cal L}
\end{eqnarray}
where ${\bar{p}}^{2}=p^{2}/k^{2}$. 
In the limits of ${\bar{p}^{2}} \to 0$ and ${\bar{p}^{2}}\to \infty$, ${\cal L}(p)$
behaves as 
\begin{eqnarray}
{\cal L}(p) \sim \frac{e^2 k^2}{2 \pi^2} \times
\begin{cases}
 {3}/{8}- {\bar{p}^{2}}/12 & ~~~~~{\bar{p}^{2}} \to 0 \\
 1/{\bar{p}^{2}}  & ~~~~~{\bar{p}^{2}} \to \infty ~. 
\end{cases}
\end{eqnarray}
The derivative expansion would give us ${\cal L}(p) \sim {3 e^2
k^2}/{16 \pi^2}$, the first term on the first line.  We also note
that ${\cal L}(p) \sim {e^2 k^4}/{2 \pi^2 p^2}$ for any non-zero $p^2$ 
in the limit of $k^2 \rightarrow 0$~.

%%%%%%%%%%%%%%%%%%%%%%%%%%%%

\section{Conclusions}\label{sec:conclusion}

In the functional renormalisation group approach to gauge theories,
gauge symmetry is encoded in the modified Ward-Takahashi identity. A
truncation to the effective action has to satisfy this mWTI. For the
Wilsonian effective action these identities can be cast into a simple
form of an unbroken symmetry identity, see
\cite{Igarashi:2001mf,Sonoda:2007dj,Igarashi:2008bb,Igarashi:2009tj}.
For the 1PI effective action the mWTI has a less convenient form due
to the presence of the full field-dependent propagator. In turn, the
flow equation for the 1PI effective action has a remarkable numerical
stability that originates in the dependence on the full
field-dependent propagator. As discussed in the introduction, this
suggests a combined use of the mWTI for the Wilsonian effective
action and the flow equation of the 1PI effective action.

In the present work we have put forward this approach for QED with a
massless electron. We have partially solved the mWTI for the Wilsonian
effective action. This solution inevitably leads to momentum
dependence couplings, or form factors.  For the sake of simplicity, we
introduced form factors only to the photon two point function and four
fermi interactions, while we ignored those in the fermion two point
function and the gauge interaction.  Even if we included the latter,
however, the generic structure observed in this paper would be the
same: higher order correlation functions would have been still
determined from lower order ones. Our solutions to the relations are
different from the one obtained by an approximation in the spirit of
derivative expansion. Finally, the related truncation for the 1PI
effective action has been calculated via the Legendre transformation,
also featuring momentum-dependent vertices. 

Note also that the present approach is easily extended to non-Abelian
gauge theories. There, in Landau gauge a BRST-consistent solution of
the flow equation is at the root of the dynamical generation of the
gluonic mass gap, see \cite{Mitter:2014wpa,CFMPS}. We hope to report on such
an extension in near future.

In summary, in the present work we have used the mWTIs to relate
couplings in the truncated Wilsonian and 1PI effective action, working
out Part (i) and (ii) of the programme put forward in the
introduction, see page two. In particular, we found that a particular
linear combination of the four fermi couplings can be written in terms
of the photon two point function as in \eq{mom.depend.relation}. It is
left to solve the flow equation for the remaining vertex
functions. This also allows to resolve the question, whether the above
WT relations are compatible with the solution of the flow equation in
the present truncation. Such a compatibility is at the root of the
overall consistency of the present approximation. This discussion, and
the solution of the gauge symmetry-consistent flows, is deferred to a
forthcoming paper.

\section*{Acknowledgment}
Y.~Igarashi and K.~Itoh are grateful to J.-I. Sumi for stimulating
discussions.  The work by Y.~Igarashi and K.~Itoh had been supported in
part by the Grants-in-Aid for Scientific Research nos. R2209 and
22540270 from Japan Society for the Promotion of Science. This work is
supported by EMMI and by ERC-AdG-290623.

\appendix

\section{Flow equation}\label{app:flow} 

Here and in the following appendices, the notation before the
renormalisation \eq{eq:gg0QED} is used so that we observe clearly how
various quantities are renormalised from the original definition of the
Wilson action to be given in \eq{int of Wilson action} and \eq{Wilson}.

In order to define the Wilson action, we introduce IR fields $\phi^{A}$
 in addition to the original fields $\varphi^{A}$ and rewrite the
 generating functional as
\begin{widetext}
\begin{align}\nn 
&\hspace{-.4cm} {\cal Z}_{\varphi}[J] = \int {\cal D} \varphi 
\exp\left(-{\cal S}[\varphi]+ J \cdot \varphi
       \right ) \\[2ex] \nn 
=&\, \int {\cal D} \varphi {\cal D} \phi \exp\left[-{\cal S}[\varphi] 
+ J \cdot \varphi-\frac{1}{2}
\Bigl(\phi - K \varphi -  J(1-K)D^{-1}\Bigr) 
\cdot
\frac{D}{K(1-K)} \cdot 
\Bigl(\phi - K \varphi - (-)^{\epsilon(J)}D^{-1}(1-K)J\Bigr)\right]\\[2ex] 
=& \, N[J] \int {\cal D} \phi 
\exp \biggl[ - \frac{1}{2} \phi \cdot ~K^{-1} D \cdot \phi - S_{I,k}[\phi]
+ J\cdot K^{-1} \phi
\biggr] \,,
\label{part-func1}\end{align}
\end{widetext} 
where interactions of $\phi$ are generated as  
\begin{align}\label{int of Wilson action}
&\hspace{-.6cm} \exp [- S_{I,k}[\phi]] \\[2ex] 
= &\int {\cal D} \chi \exp \biggl[-\frac{1}{2} 
\chi \cdot (1-K)^{-1} D \cdot \chi 
-
{\cal S}_{I}[\phi + \chi] \biggr]~. 
\nn
\end{align}
On the second line of \eq{part-func1}, a Gau\ss ian integral over
$\phi$ is inserted into the partition function.  
Changing the order of integrals over the fields $\chi= \varphi - \phi$
and $\phi$, we define the Wilson action,  
\begin{eqnarray}
S_k[\phi] = \frac{1}{2} \phi \cdot ~K^{-1} D \cdot \phi - S_{I,k}[\phi]~.
\label{Wilson}
\end{eqnarray}
The original fields $\varphi^{A}$ are decomposed into
the IR fields $\phi^{A}$ with propagator $K D^{-1}$ and the UV fields
$\chi^{A}$ with propagator $(1-K) D^{-1}$.

At this stage, we introduce $Z$ factors for IR fields and their source
terms by rescaling $\phi \to Z_\phi^{1/2}\phi$, $J_\phi \to
Z_\phi^{-1/2} J_\phi$. The Wilson action takes the form
\begin{align}
 S_{k}[Z^{1/2}\phi]
  = \frac{1}{2} \phi^{A} Z^{A}~(K^{A})^{-1}~D_{AB} \phi^{B} + 
S_{I,k}[Z^{1/2}\phi]\,.
\label{Wilsonian1}
\end{align}
The partition function for the Wilson action,
\begin{eqnarray}
 Z_{\phi}[J]
  =  \int {\cal D} \phi \exp \Bigl[-S_{k}[Z^{1/2}\phi] 
+ J \cdot K^{-1} \phi\Bigr]~,
\label{Wilsonian3}
\end{eqnarray}
is related to that for the original one as  
\begin{eqnarray}
{\cal Z}_{\varphi}[J] &=& N[Z^{-1/2}J] Z_{\phi}[J]~,
\label{part-fun2}
\end{eqnarray}
where the normalisation factor is given by 
\begin{align}\label{NJ}
& \hspace{-.6cm}N[Z^{-1/2}J]\\[2ex]  
=& \exp \left[- (-)^{\epsilon(J_{A})} \frac{1}{2}J_{A} 
\left(\frac{1-K}{Z~K}\right)^{A}
 \left(D^{-1}\right)^{AB} J_{B}\right]\,.
 \nn
\end{align}

The Polchinski flow equation is obtained from the requirement that 
the partition function ${\cal Z}_{\varphi}[J]$ does not depend on the cutoff 
$k = e^{t}$:  
$k \pa_{k} {\cal Z}_{\varphi}[J] = \pa_{t} {\cal Z}_{\varphi}[J] 
= \pa_{t} \left(N[Z^{-1/2}J]
Z_{\phi}[J]\right) =0$. It is straightforward to obtain
\begin{align}
 \label{PFE-totalapp}
& \pa_{t} S_{k}[Z^{1/2}\phi]
= - \phi^{A} \left(\pa_{t} \log K\right)^{A} 
\frac{\pa^{l} S_{k}}{\pa\phi^{A}}
\\[2ex] \nn 
&~~ + (-)^{\ep_{A}}
\frac{1}{2} 
\biggl[Z^{-1}\Bigl\{\pa_{t}K -\eta K(1-K)\Bigr\}D^{-1}
\biggr]^{AB}\\[2ex] 
& \hspace{1.6cm}\times \biggl[
\frac{\pa^{l} S_{k}}{\pa\phi^{B}}\frac{\pa^{r} S_{k}}{\pa\phi^{A}}
- \frac{\pa^{l} \pa^{r} S_{k}}{\pa\phi^{B} \pa\phi^{A}} 
\biggr]~,
\nn
\end{align}
where
\begin{eqnarray}
\eta_{A} = - \pa_{t} \log Z^{A}\,,
\label{anomalous-dimapp}
\end{eqnarray}
are the anomalous dimensions for $\phi^{A}$.  Since the Wilson action
is a functional of $(Z^{A})^{1/2} \phi^{A}$, we may extract
contributions of the anomalous dimensions from $S_{k}(Z^{1/2}\phi)$:
\begin{eqnarray}\label{PFE-total-Z fixedapp}
\pa_{t} S_{k}[\bar\phi]
&=& - \bar\phi^{A} \Bigl[\left(\pa_{t} \log K\right)^{A} - 
\frac{1}{2}\eta_{A}\Bigr]
\frac{\pa^{l} S_{k}}{\pa\bar\phi^{A}}
\\[2ex] \nn 
&& + (-)^{\ep_{A}}
\frac{1}{2} 
\biggl[\Bigl\{\pa_{t}K -\eta K(1-K)\Bigr\}D^{-1}
\biggr]^{AB}\\[2ex]
&& \hspace{1.6cm}\times \biggl[
\frac{\pa^{l} S_{k}}{\pa\bar \phi^{B}}\frac{\pa^{r} S_{k}}{\pa\bar \phi^{A}}
- \frac{\pa^{l} \pa^{r} S_{k}}{\pa\bar \phi^{B} \pa\bar \phi^{A}} 
\biggr]\,.
\nn
\end{eqnarray}
For the interaction part $S_{I,k}$, the flow equation reads \cite{Igarashi:2009tj}\cite{Sonoda01102015}
\begin{align}\label{PFE-intapp} 
  \pa_{t} S_{I,k}[\bar\phi] =&\,  \frac{1}{2}\bar\phi^{A} \eta^{A} D_{AB}~ \bar \phi^{B}
  \\[2ex]\nn 
  &\,+ \bar \phi^{A}\Bigl[\frac{1}{2}\eta - \eta (1-K)\Bigr]^{A} 
  \frac{\pa^{l} S_{I,k}}{\pa\bar \phi^{A}}
  \\[2ex] \nn 
  &\, + (-)^{\ep_{A}}
  \frac{1}{2} 
  \biggl[\Bigl\{\pa_{t}K -\eta K(1-K)\Bigr\}D^{-1}
  \biggr]^{AB}\\[2ex] 
  &\,\hspace{1.6cm}\times \biggl[
  \frac{\pa^{l} S_{I,k}}{\pa\bar\phi^{B}}
  \frac{\pa^{r} S_{I,k}}{\pa\bar\phi^{A}}
  - \frac{\pa^{l} \pa^{r} S_{I,k}}{\pa\bar\phi^{B} \pa\bar\phi^{A}} 
  \biggr]\,.
 \nn
\end{align}

\section{Derivation of the WT identity}\label{app:WTI}

We consider the realisation of gauge (BRST) symmetry for the Wilson
action. It is the WT operator that signals the presence of symmetry. We
will show that it takes the form 
\begin{eqnarray}
 \Sigma_{k}[\phi] = 
\frac{\pa^{r} S_{k}}{\pa \phi^{A}} 
\delta\phi^{A} - (-)^{\ep_{A}}
\frac{\pa^{l} \delta\phi^{A}}{\pa \phi^{A}}
~,
\label{IR-WT-op}
\end{eqnarray}
where $\delta\phi^{A}$ denote the BRST transformations for $\phi^{A}$.
The first and the second terms in \eq{IR-WT-op} are the changes of
the Wilson action and the path integral measure under the
transformation, respectively.
When the condition 
\begin{eqnarray}
\Sigma_{k}[\phi] =0 \,,
\label{IR-WT}
\end{eqnarray}
holds, we have the symmetry at the quantum level. The path integration
over the high momentum modes produces the corrections to the gauge or
BRST transformation at the scale $k$.

We now derive the WT operator (\ref{IR-WT-op}) and construct the BRST
transformations for the IR fields, $\delta \phi^{A}$: starting from the
symmetry of the UV action ${\cal S}[\varphi]$, we will find its
modification due to the presence of the cutoff.

Consider a change of variables, $\varphi^{A} \rightarrow \varphi^{A\prime} =
\varphi^{A} + \delta\varphi^{A} \lambda$, under the BRST transformation 
\begin{eqnarray}
\delta\varphi^{A} = R^{A}[\varphi]~.
\label{BRS-varphi}
\end{eqnarray}
It induces a change of the UV action
\begin{eqnarray}
{\cal S}[\varphi] \rightarrow {\cal S}[\varphi] + 
\frac{\pa^{r} {\cal S}}{\pa \varphi^{A}} \delta\varphi^{A} \lambda~,
\label{change-action}
\end{eqnarray}
as well as a change of the functional measure
\begin{eqnarray}
{\cal D}\varphi \rightarrow {\cal D}\varphi \left(1 + (-)^{A}
\frac{\pa^{l} \delta\varphi^{A}}{\pa\varphi^{A}} \lambda
\right)~.
\label{Jacobian}
\end{eqnarray}
The invariance of the functional integral
\begin{eqnarray}
  && {\cal Z}_{\varphi}[J] = \int {\cal D} \varphi' 
  \exp(-{\cal S}[\varphi'] + J\cdot \varphi')= {{\cal Z}_{\varphi}}[J]\nn \\[2ex] 
&&\hspace{.2cm}  + \int {\cal D} \varphi~
  \left[- \frac{\pa^{r} {{\cal S}}}{\pa \varphi^{A}} \delta \varphi^{A} +(-)^{A} 
    \frac{\pa^{l} }{\pa \varphi^{A}} \delta \varphi^{A} +  J \cdot \delta\varphi
  \right]\lambda \nn\\[2ex] 
  &&~~~~~~~~~~~~~~~ \times \exp\left(-{{\cal S}}[\varphi]+J \cdot \varphi
  \right)\,,
\label{inv.Z}
\end{eqnarray}
leads to a relation
\begin{eqnarray}\nn
  && \hspace{-1cm}\int {\cal D} \varphi~ {{\Sigma[\varphi]}} 
  \exp\left(-{{\cal S}}[\varphi]+J \cdot \varphi\right)\\[2ex] 
  =&& \int {\cal D} \varphi~
  J_{A}R^{A}[\varphi] \exp\left(-{{\cal S}}[\varphi]+J \cdot \varphi\right)\nn\\[2ex] 
  =&& J_{A}R^{A}\left[\pa^{l}/{\pa J}\right] {{\cal Z}_{\varphi}}[J]~.
\label{JR}
\end{eqnarray}
Here, 
\begin{eqnarray}
\Sigma[\varphi] = \frac{\pa^{r} {\cal S}}{\pa \varphi^{A}} 
\delta\varphi^{A} - (-)^{\ep_{A}}
\frac{\pa^{l} \delta\varphi^{A}}{\pa \varphi^{A}}\,,
\label{UV-WT-op}
\end{eqnarray}
denotes the WT operator for the UV action ${\cal S}$.  Using the
relation between partition functions ${\cal Z}_{\varphi}[J] =
N[Z^{-1/2}J] Z_{\phi}[J]$, we obtain the WT operator for the Wilson
action from the following calculation. 
\begin{align}
&J_{A}R^{A}\left[\pa^{l}/{\pa J}\right] {{\cal Z}_{\varphi}}[J]= 
J_{A}R^{A}\left[\pa^{l}/{\pa J}\right]  N[Z^{-1/2}J] Z_{\phi}[J]\nn\\[2ex]
& = N[Z^{-1/2}J] \int~{\cal D} \phi~ \Sigma_{k}[\phi] 
\exp\left(-S_{k}[\phi]+J \cdot K^{-1} \phi\right)~.
\label{WT3}
\end{align}
In the second expression in \eq{WT3}, derivatives
$R^{A}\left[\pa^{l}/{\pa J}\right]$ act on $N[Z^{-1/2}J]$ as well as the
partition function.  This generates the modified BRST transformation
$\delta \phi^{A}$.  It is easy to confirm that $\Sigma_k$ in
\eq{WT3} agrees with the expression in \eq{IR-WT-op}.

\subsection{Composite operators} \label{app:composite}

The WT operator $\Sigma_{k}$ is characterised by 
composite operators. We summarise some results on them \cite{Igarashi:2009tj}. 
An operator ${\cal O}_{k}[\phi]$ is
called a composite operator if it fulfills a RG flow equation
\begin{widetext} 
\begin{eqnarray}
\pa_{t} {\cal O}_{k}[\phi] &=& 
 - \phi^{A} \left(\pa_{t} \log K \right)^{A}
 \frac{\pa^l {\cal O}_{k}}{\pa \phi^A}
 +  \frac{\pa^r S_{k}}{\pa\phi^A}
 \biggl[Z^{-1}\Bigl\{\pa_{t} K -  \eta K(1-K)\Bigr\}D^{-1}
 \biggr]^{AC}\frac{\pa^l {\cal O}_{k}}{\pa \phi^C}\nn\\
 && - \frac{1}{2}
 \biggl[Z^{-1}\Bigl\{\pa_{t} K -  \eta K(1-K)\Bigr\}D^{-1}
 \biggr]^{AC} 
\frac{\pa^l \pa^{l} {\cal O}_{k}}{\pa \phi^C \pa \phi^A}\,.
\label{composite-flow}
\end{eqnarray}
\end{widetext}
This flow equation takes the same form as a variation of 
the Polchinski equation (\ref{PFE-totalapp}) for 
an infinitesimal deformation of the Wilson action, $\Delta S_{k}$.  
The composite fields $[\phi]_{k}$ defined as
\begin{eqnarray}\nonumber 
  \left[\phi^{A}\right]_{k} & \equiv & 
  \left(K^{-1} \phi\right)^{A} -  
  \left(Z^{-1}G^{(0)}\right)^{AB} 
  \frac{\pa^{l} S_{k}}{\pa \phi^{B}}\\[2ex] 
  & = & \phi^{A} -  \left(Z^{-1}G^{(0)}\right)^{AB} 
  \frac{\pa^{l} S_{I,k}}{\pa \phi^{B}}\,,
\label{composite}
\end{eqnarray} 
play an important role in constructing $\Sigma_{k}$. Note that
$\left[\phi^{A}\right]_{k}$ equal the full mean fields $\Phi$ in the
1PI language defined in \eq{field-relation}.  Then,
\eq{delta-phi-composite0} is obtained from \eq{linear-BRS} simply by
multiplying the function $K$ and replacing classical fields with the
corresponding composite operators \cite{Igarashi:2009tj}.  In addition
to $\left[\phi^{A}\right]_{k}$, $K^{A} \pa S_{k}/\pa \phi^{A}$ and the
WT operator $\Sigma_{k}$ itself are composite operators. As a result,
once the identity $\Sigma_{\hat k}[\phi] =0$ is shown at some scale
$\hat k$, $\Sigma_{k}[\phi]$ vanishes at any scale $k$. Therefore, the
WT identity can be used to define a gauge invariant subspace in the
theory space.

\subsection{The WT identity for the Wilson action for QED}

We will describe the construction of the quantum BRST transformation and
the WT operator for QED with fields $\phi^{A} =
(a_{\mu},~\bar\psi_{\hat\alpha},~\psi_{\alpha},~c,~\bar{c})$.
Though we use the same notations for the component fields as for $\bar
\phi$, here they all represent the unrenormalised fields. The kinetic
part of the Wilson at the scale $k$ is given by
\begin{widetext} 
\begin{eqnarray}
  S_{0, k}
&=& \int_p K^{-1}(p)\biggl[\frac{Z_{3}}{2}a_{\mu}(-p)~p^{2}
\left\{\delta_{\mu\nu}- \left(1- {\xi^{-1}} \right) 
\frac{p_{\mu}p_{\nu}}{p^2}\right\} a_{\nu}(p) 
+ {\bar c}(-p)ip^{2}c(p) \biggr] + \int_{p} K^{-1}(p)Z_{2} 
{\bar \psi}(-p) \Slash{p} \psi(p)~\nn\\[2ex]
&=& \frac{1}{2} (K^{A})^{-1}Z_{A}\phi^{A} D_{AB} \phi^{B}\,,
\label{Wilson5}
\end{eqnarray}
\end{widetext} 
where $Z_{2}$ and $Z_{3}$ are the renormalisation constants of the
fermion and photon fields respectively.  For simplicity, the fermion
is chosen to be massless and we have the chiral symmetry. The matrix
$D_{AB}$
\begin{align} 
D_{AB}(p)
\equiv 
  \left(
    \begin{array}{ccc}
      \left(D_{G}\right)_{\mu\nu}(p) & 0  & 0  \\
      0  & 0  & \left(D_{F}\right)_{{\hat \al} {\beta}}(p) \\
      0  & \left(D_{F}\right)^{T}_{{\al} {\hat \beta}}(p) & 0
    \end{array}
  \right)  \,,
\end{align}
has the components, 
\begin{align}
\left(D_{G}\right)_{\mu\nu}(p) =  p^2 \left(P^{T}_{\mu\nu} 
+ \xi^{-1} P^{L}_{\mu\nu}\right)\,,
\label{Dmunu}\end{align} 
with 
\begin{align}
P^{T}_{\mu\nu} = \delta_{\mu\nu} - \frac{p_{\mu}p_{\nu}}{p^2}\,,\qquad 
P^{L}_{\mu\nu} = \frac{p_{\mu}p_{\nu}}{p^2}\,,
\end{align} 
and 
\begin{align}
\left(D_{F}\right)_{\hat\al\beta}(p) = \left(\Slash{p}\right)_{\hat\al\beta}\,,\quad 
\left(D_{F}\right)^{T}_{\al\hat\beta}(p) = \left(\Slash{p}^{T}\right)_{\al\hat\beta}\,.
\label{Dalbe}
\end{align}
The high energy propagators are expressed as 
\begin{eqnarray}
&&\left(G^{(0)}\right)^{AB}(p) =(1-K) \left(D^{-1}\right)^{AB}(p)
 \\
&&\equiv 
  \left(
    \begin{array}{ccc}
      \left(G_{G}^{(0)}\right)_{\mu\nu}(p) & 0  & 0  \\
      0  &  0  & \left(G_{F}^{(0)}\right)^T_{{\hat \al} {\beta}}(p) \\
      0  & \left(G_{F}^{(0)}\right)_{{\al} {\hat \beta}}(p) & 0 
    \end{array}
		      \right)\,,
  \nn
\end{eqnarray}
where
\begin{eqnarray}\nn 
\left(G_{G}^{(0)}\right)_{\mu\nu}(p) &=& \frac{1-K}{~p^2} 
\left(P^{T}_{\mu\nu} 
+ {\xi} P^{L}_{\mu\nu}\right)~, 
\\[2ex] \nn 
\left(G_{F}^{(0)}\right)_{\al\hat\beta}(p) &=& (1-K) 
\left(\frac{1}{\Slash{p}}\right)_{\al\hat\beta}\,,\\[2ex] 
\left(G_{F}^{(0)}\right)^T_{{\hat \al} {\beta}}(p)&=&
(1-K) 
\left(\frac{1}{\Slash{p}^{T}}\right)_{{\hat \al} {\beta}}~.
\end{eqnarray}
By starting from \eq{classical-BRS}, the quantum BRST transformation
for QED is obtained by multiplying $K(p)$ and replacing the fields by
their composite operators on the r.h.s.  This procedure is based on
the observation that $K^{-1} \delta \phi^A$ are composite operators
\cite{Igarashi:2009tj}.
\begin{eqnarray}
K^{-1}\delta a_{\mu}(p) &=& -i p_{\mu} c(p)~,\nn\\[2ex]
K^{-1}\delta \psi(p) &=& -i~e_{0}~\int_{q} [\psi(q)]_k c(p-q)~,\nn\\[2ex]
K^{-1}\delta {\bar \psi}(-p) &=& i~e_{0} \int_{q}  [{\bar \psi}(-q)]_k c(q-p)~,\nn\\[2ex]
K^{-1}\delta {\bar c}(p) &=& \xi_{0}^{-1} p_{\mu}~[a_{\mu}(p)]_k~. 
\label{q-BRS}
\end{eqnarray}
Here $[\phi^A]_k$ are composite fields given in
\eq{composite}. Written in terms of $\bar\phi$ and $\bar \Phi$, we
obtain \eq{quantum-BRS}. By substituting \eq{q-BRS} into \eq{IR-WT-op}, we
obtain the WT operator for QED,
\begin{widetext}
\begin{eqnarray} 
{\hspace{-1cm}}\Sigma_{k}[\phi]
 &=& \int_{p}\biggl[
 K(p)\frac{\pa S_{k}}{\pa a_{\mu}(p)}(-ip_{\mu})c(p) +
K(p)\frac{\pa^{r} S_{k}}{\pa {\bar c}(p)} \Bigl\{Z_{3}
\xi^{-1} p_{\mu} K^{-1}(p)a_{\mu}(p) -  
\frac{(1-K(p))}{p^{2}} p_{\mu} \frac{\pa S_{k}}{\pa a_{\mu}(-p)}
\Bigr\}
\biggr]\nn\\[2ex] 
&~&-i e_{0} \int_{p,q} \biggl[\frac{\pa^{r} S_{k}}{\pa
 \psi_{\al}(q)}\frac{K(q)}{K(p)}
\psi_{\al}(p)
 -\frac{K(p)}{K(q)}{\bar \psi}_{\hat\al}(-q)
\frac{\pa^{l} S_{k}}{\pa {\bar \psi}_{\hat\al}(-p)} \biggr]c(q-p)\nn\\[2ex] 
 &~& -i e_{0} \int_{p,q} U_{\beta\hat\al}(-q,p) \biggl[
\frac{\pa^{l} S_{k}}{\pa {\bar \psi}_{\hat\al}(-p)} 
\frac{\pa^{r} S_{k}}{\pa \psi_{\beta}(q)}
- \frac{\pa^{l}\pa^{r} S_{k}}{\pa {\bar \psi}_{\hat\al}(-p)\pa 
\psi_{\beta}(q)}
\biggr]c(q-p)\nn\\[2ex] 
&& = \int_{p}\biggl[
\frac{\pa S_{k}}{\pa a_{\mu}(p)}(-ip_{\mu})c(p) + Z_{3}
\frac{\pa^{r} S_{k}}{\pa {\bar c}(p)} \xi^{-1} p_{\mu}a_{\mu}(p) 
\biggr] \label{WT-QED1app} \\[2ex] 
&~&-i~e_{0} \int_{p,q} \biggl[\frac{\pa^{r} S_{k}}{\pa
 \psi_{\al}(q)}\frac{K(q)}{K(p)}
\psi_{\al}(p)
 -\frac{K(p)}{K(q)}{\bar \psi}_{\hat\al}(-q)
\frac{\pa^{l} S_{k}}{\pa {\bar \psi}_{\hat\al}(-p)} \biggr]c(q-p)\nn\\[2ex] 
&~& -i~e_{0} \int_{p,q} U_{\beta\hat\al}(-q,p) \biggl[
\frac{\pa^{l} S_{k}}{\pa {\bar \psi}_{\hat\al}(-p)} 
\frac{\pa^{r} S_{k}}{\pa \psi_{\beta}(q)}
- \frac{\pa^{l}\pa^{r} S_{k}}{\pa {\bar \psi}_{\hat\al}(-p)\pa 
\psi_{\beta}(q)}
\biggr]c(q-p)~,
\nn\\[2ex] 
 &=&
  \int_{p}\frac{\pa S_{I,k}}{\pa a_{\mu}(p)}(-ip_{\mu})c(p)
 + i~e_{0} 
Z_{2}\int_{p,q}\Bigl[\bar\psi(-q)(\Slash{p} -\Slash{q})\psi(p)
\Bigr]c(q-p)\nn\\[2ex] 
&~&-i~e_{0} \int_{p,q} \biggl[\frac{\pa^{r} S_{I,k}}{\pa
 \psi_{\al}(q)}\frac{K(q)}{K(p)}
\psi_{\al}(p)
 -\frac{K(p)}{K(q)}{\bar \psi}_{\hat\al}(-q)
\frac{\pa^{l} S_{I,k}}{\pa {\bar \psi}_{\hat\al}(-p)} \biggr]c(q-p)\nn\\[2ex] 
&~& +  i~e_{0} Z_{2}\int_{p,q}
\biggl[\frac{\pa^{r} S_{I,k}}{\pa
 \psi_{\beta}(q)}U_{\beta\hat\al}(-q,p)K^{-1}(p)\Bigl(\Slash{p} \psi(p)\Bigr)
_{\hat\al} %\nn\\[2ex] 
%&~&
+ K^{-1}(q)\Bigl(\bar\psi(-q)\Slash{q}\Bigr)_{\beta}
U_{\beta\hat\al}(-q,p)\frac{\pa^{l} S_{I,k}}{\pa {\bar \psi}
_{\hat\al}(-p)} 
\biggr]c(q-p)\nn\\[2ex] 
&~& - i~e_{0}  \int_{p,q} U_{\beta\hat\al}(-q,p) \biggl[
\frac{\pa^{l} S_{I,k}}{\pa {\bar \psi}_{\hat\al}(-p)} 
\frac{\pa^{r} S_{I,k}}{\pa \psi_{\beta}(q)}
-\frac{\pa^{l}\pa^{r} S_{I,k}}{\pa {\bar \psi}_{\hat\al}(-p)\pa 
\psi_{\beta}(q)}
\biggr]c(q-p)~,
\nn
\end{eqnarray}
\end{widetext} 
where
\begin{eqnarray}
 U(-q,p)_{\beta\hat\al} 
= Z_{2}^{-1}\Bigl[K(q)
G_{F}^{(0)}(p) - K(p) 
G_{F}^{(0)}(q)\Bigr]_{\beta\hat\al}  ~.\nn\\
\label{U matrix 2app}
\end{eqnarray}
In the last expression of \eq{WT-QED1app}, the WT operator
$\Sigma_{k}[\phi]$ is expressed in terms of the interaction part of
the Wilson action, $S_{I,k}$. Since the gauge parameter $Z_{3}
\xi^{-1} = \xi_{0}^{-1}$ and gauge coupling $e_{0}$ are $k$
independent, $\Sigma_{k}[\phi]$ remains a composite operator.

Now, in order to write \eq{q-BRS} and \eq{WT-QED1app} with the
renormalised fields $\bar \phi$, we make a rescaling $a_{\mu} \to
Z_{3}^{-1/2}a_{\mu},~\bar\psi \to Z_{2}^{-1/2} \bar\psi,~\psi \to
Z_{2}^{-1/2} \psi,~c \to Z_{e} c,~ \bar{c} \to Z_{e}^{-1} \bar{c}$,
where $Z_{e} =e/e_{0}$.  Then, we obtain the BRST transformations
\eq{quantum-BRS} and the WT identity \eq{WT-QED1}.

\section{Wilson effective action for QED} \label{app:Wilson} 
In Section~\ref{sec:GI+SI} we have discussed the approximation to the
1PI effective action and the Wilson effective action used in the
present work. With all momentum dependence and Lorentz indices the
schematic expression \eq{eq:GIschem} is given by
\begin{widetext} 
\begin{eqnarray}\nonumber 
  \Gamma_{I,k}[\bar\Phi]&=& \frac{Z_{3}}{2} \int_p A_{\mu}(-p)A_{\nu}(p) 
  h^{(aa)}_{\mu\nu}(p)+ Z_{2}
  \int_p \bar\Psi_{\hat\alpha}(-p) \Psi_{\alpha}(p) 
  h^{(\bar\psi\psi)}_{\hat\alpha\alpha}(p)\nn\\[2ex]  && 
  -e Z_{2} Z_{3}^{1/2}\!\int_{p_{1},p_{2}}\!\!\!\!\!\!\bar\Psi_{\hat\alpha}(p_{1})  
  A_{\mu}(-(p_1+p_2)) \Psi_{\alpha}(p_{2})
  h^{(\bar\psi a\psi)}_{\hat\alpha\alpha,\mu}(p_{1},p_{2}) \nn\\[2ex] 
  && + Z_{2}^{2}
  \int_{p_{1},p_2,p_{3}} \bar\Psi_{\hat\alpha}(p_{1})  
  \Psi_{\alpha}(p_{2}) 
  \bar\Psi_{\hat\beta}(p_{3})  
  \Psi_{\beta}(-(p_1+p_2+p_3))  h^{(\bar\psi\psi\bar\psi\psi)}_{\hat\alpha\alpha\hat\beta\beta}
  (p_{1},p_{2},p_{3}) \,. 
\label{eq:GI-QED}\end{eqnarray}
Correspondingly, $S_{I,k}$ is given by
\begin{eqnarray} S_{I,k}[\phi]& = & \frac{Z_{3}}{2} \int_p
  a_{\mu}(-p)a_{\nu}(p)
  \Bigl[\bar{G}_{G}(p)\tilde{h}^{(aa)}(p)\Bigr]_{\mu\nu} + Z_{2}
  \int_p \bar\psi_{\hat\alpha}(-p) \psi_{\alpha}(p)
  \Bigl[\bar{G}_{F}(p) \tilde{h}^{(\bar\psi\psi)}(p)\Bigr]
  _{\hat\alpha\alpha} \nn \\[2ex]  && -e
  Z_{2}Z_{3}^{1/2}\int_{p_{1},p_{2}} \bar\psi_{\hat\alpha}(p_{1})
  a_{\mu}((-(p_1+p_2)) \psi_{\alpha}(p_{2})
  \Bigl[\prod_{i=1}^{2}\bigl\{\bar{G}_{F}(p_{i})\bigr\}
  \bar{G}_{G}(-(p_1+p_2)) \tilde{h}^{(\bar\psi
    a\psi)}(p_{1},p_{2})\Bigr]_{\hat\alpha\alpha,\mu} \nn \\[2ex]  &&
  + Z_{2}^{2} \int_{p_{1},p_2,p_3} \bar\psi_{\hat\alpha}(p_{1})
  \psi_{\alpha}(p_{2}) \bar\psi_{\hat\beta}(p_{3})
  \psi_{\beta}((-(p_1+p_2+p_3))
  \Bigl[\prod_{i=1}^{4}\bigl\{\bar{G}_{F}(p_{i})\bigr\}
  \tilde{h}^{(\bar\psi\psi\bar\psi\psi)}(p_{1},p_{2},p_{3})
  \Bigr]_{\hat\alpha\alpha\hat\beta\beta} \,.
\label{SI-QED} 
\end{eqnarray}
In this work we further reduce the general tensor structure of the
interaction coefficients $h^{(\bar\psi \psi)}$, $h^{(\bar\psi a\psi)}$
and $h^{(\bar\psi\psi\bar\psi \psi)}$, see \eq{hH2}-\eq{hH4},
\eq{eq:4fermi1} and \eq{Gamma_I}. This leads to

  \begin{eqnarray} 
    S_{I,k}& = &\int_{p}~\biggl[\frac{Z_{3}}{2}
    a_{\mu}(-p)
    \Bigl[
    {\bar G}_G(p){h^{(aa)}(p)}
    \Bigr]_{\mu\nu}
    a_{\nu}(p) +  Z_{2} \tilde\sigma(p)\sigma(p)
    {\bar \psi}(-p)\Slash{p} \psi(p)\biggr]\nn\\[2ex] 
    &&-  e Z_{2}Z_{3}^{1/2}\int_{p,q} {h_{e}(-p,q)}
    \tilde{\sigma}(p)
    \bar\psi(-p)
    \gamma_{\mu}\tilde{\sigma}(q)\psi(q)
    \bigl({\bar G}_G\bigr)_{\mu \nu}(p-q)
    ~a_{\nu}(p-q)\nn\\[2ex]
    && + \frac{Z_{2}^{2}}{2} \int_{p_{1},\cdots,p_{3}}\biggl[\prod_{i=1}^{4} \tilde\sigma(p_{i})\biggr]
\label{SI}\biggl[ 
\frac{ h_{S}(s,t,u)}{k^{2}}
\Bigl\{\left(
{\bar\psi}(p_{1})\psi(p_{2})\right)\left({\bar\psi}(p_{3})\psi(p_{4})\right)
 - \left({\bar\psi}(p_{1})\gamma_{5}\psi(p_{2})\right)\left({\bar\psi}(p_{3})
\gamma_{5}\psi(p_{4})\right)\Bigr\}\nn\\[2ex]
&& \hspace{2.5cm}+ \frac{h_{V}(s,t,u)}{k^{2}}\Bigl\{\left(
{\bar\psi}(p_{1})\gamma_{\mu}\psi(p_{2})\right)
\left({\bar\psi}(p_{3})\gamma_{\mu}
\psi(p_{4})\right)
+ \left({\bar\psi}(p_{1})\gamma_{5}\gamma_{\mu}\psi(p_{2})\right)
\left({\bar\psi}(p_{3})
\gamma_{5}\gamma_{\mu}\psi(p_{4})\right)\Bigr\}\nn\\[2ex]
&& \hspace{2.5cm}- e^{2} {h_{e}(p_{1},p_{2})}{h_{e}(p_{3},p_{4})}
\left(
{\bar\psi}(p_{1})\gamma_{\mu}\psi(p_{2})\right)
\left({\bar\psi}(p_{3})\gamma_{\nu}
\psi(p_{4})\right)
\left(G_G\right)_{\mu\nu}(p_{1}+p_{2})
\biggr]
\nn\\[2ex]
&& + \frac{Z_{2}^{2}}{2k^{4}}\int_{p_{1},\cdots,p_{3}}
\biggl[\prod_{i=1}^{4} \tilde\sigma(p_{i})\biggr]
h_{V'}(s,t,u)(p_{1}+p_{4})_{\mu}(p_{2}+p_{3})_{\nu}\nn\\[2ex]
&&\hspace{2.5cm}\times \biggl[ 
\left({\bar\psi}(p_{1})\gamma_{\mu}\psi(p_{2})\right)
\left({\bar\psi}(p_{3})\gamma_{\nu}
\psi(p_{4})\right) 
+ \left({\bar\psi}(p_{1})\gamma_{5}\gamma_{\mu}\psi(p_{2})
\right)
\left({\bar\psi}(p_{3})\gamma_{5}\gamma_{\nu}
\psi(p_{4})\right)
\biggr]~. 
\label{eq:SIk}
  \end{eqnarray}
\end{widetext}
with $p_4 = -(p_1+p_2+p_3)$ and  
\begin{eqnarray}
\tilde\sigma(p)=\frac{1}{1 +(1-K(p))\sigma(p)}~.
\label{sigmatilde}
\end{eqnarray}
\ 
\eject 
\ 
\section{Derivation of \eq{2nd relation2} and \eq{cal L}}\label{app:der2}

 Before using our ansatz \eq{assumption}, the second WT relation 
takes the following form:
\begin{widetext} 
\begin{eqnarray}
&& e_{0} Z_{2}\Bigl[\Slash{p}\bigl(1 + \sigma(p)\bigr) 
-\Slash{q}\bigl(1 + \sigma(q)\bigr)\Bigr]  
- e Z_{2}Z_{3}^{1/2}
{h_{e}(-q,p)} \gamma_{\mu} (p-q)_{\nu}
  \bigl({\bar G}_G \bigr)_{\mu \nu}(p-q)
\nn\\[2ex] 
&& - e_{0}  
\frac{Z_{2}^{2}}{k^{2}}
\int_{l} \biggl[ 2h_{S}\bigl(l^{2},(p+q+l)^{2},(p-q)^{2}\bigr)
U(-q-l,p+k)\nn 
\\ [2ex] 
&& 
~~~~~~+ 2h_{V}\bigl(l^{2},(p+q+l)^{2},(p-q)^{2}\bigr) 
~\gamma_{\mu}U(-q-l,p+l)\gamma_{\mu}\nn\\[2ex] 
&& ~~~~~
- h_{V}\bigl((p-q)^{2},(p+q+l)^{2},l^{2}\bigr)
~{\rm Tr}[U(-q-l,p+l)\gamma_{\mu}]\gamma_{\mu}
\biggr]{\tilde\sigma}(p+l){\tilde\sigma}(q+l)
\nn
\end{eqnarray}
\begin{eqnarray}
&& + e_{0} 
\frac{Z_{2}^{2}}{k^{4}} \int_{l} \biggl[ 2h_{V'}
\bigl(l^{2},(p+q+l)^{2},(p-q)^{2}\bigr)
(\Slash{p} - \Slash{q})U(-q-l,p+l)(\Slash{p} - \Slash{q})\nn\\[2ex] 
&& ~~~~~ - h_{V'}\bigl((p-q)^{2},(p+q+l)^{2},l^{2}\bigr)~ 
\Slash{l} {\rm Tr}\Bigl\{U(-q-l,p+l)\Slash{l}\Bigr\} 
\biggr]{\tilde\sigma}(p+l){\tilde\sigma}(q+l)\nn\\[2ex] 
&& ~~~~~ - e_{0} e^{2}Z_{2}^{2} Z_{3}
\int_{l}\tilde{\sigma}(l)\tilde{\sigma}(p-q-l)\biggl[h_{e}(-q,p) 
h_{e}(-l,-p+q+l)\nn\\[2ex] 
&& ~~~~\times \gamma_{\mu} 
{\rm Tr}[U(p-q-l,l)\gamma_{\nu}](G_{G})_{\mu\nu}(p-q) \nn\\[2ex] 
&& ~~~~ - h_{e}(-l,p)h_{e}(-q,-p+q+l)
\gamma_{\mu} U(p-q-l,l)\gamma_{\nu} 
\left(G_G\right)_{\mu\nu}(p-l)\biggr] =0 \,. 
\label{2nd relation1}
\end{eqnarray}
In \eq{2nd relation1}, one-particle reducible contributions are
summed up to give
\begin{eqnarray} 
  && \hspace{-.5cm}e Z_{2}Z_{3}^{1/2} h_{e}(-q,p)
  \gamma_{\mu} (p-q)_{\nu}
  ({\bar G}_G)
  _{\mu \nu}(p-q)
  + e_{0} e^{2} Z_{2}^{2} h_{e}(-q,p) \left(G_{G}\right)_{\mu\nu}(p-q)\nn\\[2ex] 
&& \times  \int_{l} {\rm Tr}[U(p-q-l,l)\gamma_{\mu}]
  \gamma_{\nu} h_{e}(-l,p-q-l) \tilde{\sigma}(l)\tilde{\sigma}(p-q-l) 
\nn\\[2ex]
  && = e Z_{2}Z_{3}^{1/2}h_{e}(-q,p)
  (\Slash{p} -\Slash{q})
  \left[\frac{(p-q)^{2}}{(p-q)^{2} +\xi \left\{(1-K){\cal L}\right\}(p-q)} 
    + \frac{\xi \left\{(1-K){\cal L}\right\}(p-q)}
    {(p-q)^{2} +\xi \left\{(1-K){\cal L}\right\}(p-q)}\right] \nn\\[2ex] 
&&= e Z_{2}Z_{3}^{1/2} h_{e}(-q,p)
  (\Slash{p}-\Slash{q})\,.  
\label{sum of two terms}
\end{eqnarray}
Here we have used the extended form of the first WT relation with 
form factors $h_{e}(p,q)$ and $\sigma(p)$: 
\begin{eqnarray}
p_{\mu} {\cal L}(p) = e_{0} e Z_{2}Z_{3}^{-1/2} \int_{l}
{\rm Tr}[U(p-l,l)\gamma_{\mu}]h_{e}(-l,-p+l)
\tilde{\sigma}(l)
\tilde{\sigma}(-p+l)~.
\label{1st relation1}
\end{eqnarray}
\end{widetext}
Note that the propagators $G_{F}$ and $G_{G}$ contain the inverse of Z
factors, $U \propto Z_{2}^{-1},~G_{G} \propto Z_{3}^{-1}$.  Using the
locality assumption $\sigma(p) = 0\quad (\tilde{\sigma} =1),~h_{e}(p,q)
=1$, the relation $e= Z_{e} e_{0}$, and some trace relations such as
${\Slash{l}}{\rm Tr}\left[U(-q,p){\Slash{l}} \right] = 2
{\Slash{l}}U(-q,p){\Slash{l}} + 2 l^{2}U(-q,p)$, we obtain \eq{2nd
relation2}.

Let us consider the 
first WT relation \eq{1st relation1}, which reduces to
\begin{align}
p_{\mu}{\cal L}(p) =&\, 
e^{2} Z_{2} 
\int_{q}~{\rm Tr}\left[U(p-q,q)\gamma_{\mu}\right]\nn\\[2ex]
=&\, -4 e^{2} \int_{q}~\Bigl[K(q+p)- K(q-p)\Bigr]
q_{\mu}\frac{1-K(q)}{q^{2}}\,,
\label{calculation of L}
\end{align}
with our present assumption \eq{assumption}.  Using the Gau\ss ian
function for the cutoff function \eq{gaussian} and the integral
representation of the modified Bessel function
\begin{align} 
\int_{0}^{\pi} d \theta \exp (2\bar{p}\bar{q} \cos\theta)\sin^{2}\theta = 
\frac{\pi}{2\bar{p}\bar{q}} I_{1}(2\bar{p}\bar{q})\,,
\end{align}
we have
\begin{eqnarray}
  &&  \int_{\bar{q}} \exp (-\bar{q}^{2} + 2 \bar{p}\cdot \bar{q})\left(\frac{1 -
      \exp (-\bar{q}^{2})}{\bar{q}^{2}}
  \right) \nn\\[2ex] 
  &=& \frac{1}{8\pi^{2}\bar{p}} \int_{0}^{\infty} d\bar{q}~
  \left[\exp(-\bar{q}^{2}) - \exp(-2\bar{q}^{2})\right]I_{1}(2\bar{p}\bar{q})
  \nn\\[2ex] 
  &=& \frac{1}{16\pi^{2} \bar{p}^{2}} \left[\exp(\bar{p}^{2}) - 
    \exp(\bar{p}^{2}/2)\right]\,,
\label{mod Bessel integrals}
\end{eqnarray}
where ${\bar{p}}^{2}=p^{2}/k^{2}$ and ${\bar{q}}^{2}=q^{2}/k^{2}$.
From \eq{calculation of L} with \eq{mod Bessel integrals}, we obtain
\eq{cal L}.

%%%%%%%%%%%%%%%%%%%%%%%
%

%%%%%%%%%%%%%%%%%%%%%%%
%\bibliography{fQED}
\end{document}